# Evolutionary Design of Interpretable Spatial Descriptors for Structure-Property Relationship Discovery

Aditya Raghavan* and Sergei V. Kalinin†

[1]Department of Materials Science and Engineering, University of Tennessee, Knoxville, TN37922

## Abstract

Structure–property prediction is a classical task in machine learning, ranging from molecular activity and materials properties to local functional responses measured in microscopy. For image-based regression, two choices define the information available to the predictor: the size of the image neighborhood and the spatial resolution retained within it. A small neighborhood can omit relevant surrounding structure, whereas a large, finely sampled neighborhood can introduce redundant or irrelevant information. Here, we develop a framework for constructing spatial descriptors for prediction of local spectra and hysteresis loops. We benchmark regression models, use SHAP maps to examine their spatial attribution, and explore the effect of the patch size and compression to establish the relationship between predictive accuracy and descriptor dimensionality. We then use a genetic algorithm to optimize spatial partitions and the assignment of mean, minimum, and maximum pooling operations. Each retained region contributes an explicit measurement of the image, allowing the search to balance prediction error against the number of descriptor values. Cartesian and polar constructions provide complementary ways to organize this information. The resulting descriptors make it possible to examine where spatial detail is needed, where regional summaries are sufficient, and how these choices depend on the predictor and the measured response. The same approach can be used for molecules and solids once the genetic algorithm rules for these are established.

* araghav4@vols.utk.edu
† sergei2@utk.edu

## I. Introduction

A central task in supervised machine learning is to establish relationships between observations that are readily accessible and properties that are expensive, slow, or difficult to measure or calculate. Molecular examples include prediction of atomization energies from atomic identities and coordinates, and prediction of solubility or biological activity from molecular structure[1-3]. In inorganic materials, composition and crystal structure can serve as inputs for prediction of formation energies, electronic properties, and other functional characteristics[4-6]. In microscopies and nanoindentation, the corresponding problem is to infer local properties or stimulus-response spectra from an image of the surrounding structure. Such predictions offer a means to prioritize detailed measurements and interrogate structure–property relationships, provided their domain of validity is established.

This general structure–property discovery task has stimulated the development of datasets, algorithms, and representations. For molecules, reference calculations and curated experimental collections support prediction of energetic, physicochemical, and biological properties, as exemplified by molecular energy benchmarks and MoleculeNet[2, 3]. For crystalline solids, the Materials Project provides systematically computed structures and properties that enable comparisons across material classes[5]. Recent work on materials force fields also highlight the problem with redundancy in large training datasets, emphasizing importance of local atomic neighborhoods over whole datasets[7]. Microscopy has a less standardized and more fragmented landscape of paired structural and functional data. Nevertheless, image-to-spectrum learning, domain-structure-to-switching prediction, and autonomous spectroscopy have already demonstrated how local structural observations can be connected to measured responses[8-11]. These studies, including our earlier work, establish the scientific opportunity even in the absence of comparably broad, standardized microscopy structure–property databases.

A descriptor is a numerical representation that maps a structure or observation to the variables supplied to a prediction model. Early descriptors were often chosen from accessible measurements or established physical and chemical quantities; modern representations range from molecular fingerprints and materials feature libraries to local atomic environments and learned embeddings[1, 4, 6, 12, 13]. Availability, however, does not establish predictive relevance. Irrelevant or redundant variables can reduce generalization performance in finite datasets by increasing the estimation burden or providing opportunities to fit noise, although the effect depends on the predictor and regularization[14]. Descriptor construction also raises an associated question of interpretability: which aspects of the input support a prediction? LIME and SHAP are widely used to characterize fitted models through local surrogate explanations and additive feature attributions, respectively[15, 16]. Attention mechanisms provide learned weighting of contextual information within neural architectures[17], although attention weights alone do not establish a physical explanation.

These considerations open a complementary question: when paired structures and properties are available, how can we construct descriptors that are both optimally predictive and directly interpretable? Attribution explains the behavior of an existing predictor, but does not itself specify the smallest or most appropriate representation for a new regression task. For image-based structure–property relationships, this requires deciding which spatial neighborhood to retain, how finely to resolve it, and which measurements to extract from its different regions.

Locality provides a useful starting hypothesis, but it does not determine the optimal descriptor. Nearby structure may dominate a response, while interfaces, neighboring objects, collective interactions, or the spatial sensitivity of the instrument can introduce contributions from a wider neighborhood. Increasing patch size changes the physical context; changing sampling or compression changes how that context is represented. Larger descriptor sizes or finer resolution can supply useful information, but they can also introduce correlated or noisy variables whose value depends on training-set size and regularization. High dimensionality therefore creates a model-dependent trade-off rather between predictive power and accuracy, necessitating development of optimal descriptors for the predictive tasks.

Here, we investigate this trade-off for image-to-spectrum and image-to-hysteresis prediction. We first benchmark regressors and characterize spatial attribution patterns. We then vary patch size and uniform compression, before using a genetic algorithm to construct adaptive spatial descriptors. The search acts on the support and pooling operation of each feature, where the latter can represent the minimum, maximum, or mean intensity within an evolved region. This approach shares the compactness objective of symbolic model discovery, while optimizing a spatial measurement representation coupled to a separately fitted predictor. Cartesian and polar descriptor families allow the consequences of geometry and spatial resolution to be examined directly.

## II. Prior work

Descriptor construction encompasses three related problems: selecting variables from an existing library, constructing new combinations of those variables, and defining the support over which a measurement is made. QSAR established the practical value of structural descriptors for predicting molecular responses[1], while materials feature libraries extended systematic descriptor-based regression to composition and structure[4, 12]. Symbolic approaches seek compact combinations of primary quantities. SISSO generates a large candidate feature space through mathematical operators, applies sure independence screening, and uses sparsifying selection to identify a low-dimensional model[18]. PySR evolves symbolic expressions by mutation and recombination, together with simplification and numerical optimization of constants[19]. For regression, predictive loss is considered alongside expression complexity, producing a family of accurate, compact expressions. These approaches establish representation discovery as part of model construction, but operate primarily on mathematical expressions.

The separation between an evolved representation and a fitted regressor is also established. La Cava and Moore represent a candidate feature space as a set of programs and study evolutionary operators that promote useful, complementary features for regression[20]. Their work is an important algorithmic precedent because it evaluates features through the performance of a downstream model. It also illustrates the limitation of counting features alone, since a small descriptor can still contain strongly correlated values. Our representation restricts the search to spatial supports and regional statistics, making each feature directly identifiable on the image rather than allowing an unrestricted algebraic program.

The closest precedent identified in the reviewed literature is Rogers et al., who use genetic programming to select wavelength regions and extract informative features from hyperspectral sediment images for prediction of organic matter and porosity[21]. This is direct prior work on evolutionary image descriptors for physical-property regression. Earlier, Chion et al. developed genetic-programming spectral vegetation indices for quantitative agricultural prediction[22]. These studies introduce the general idea of evolving image-derived predictors of a continuous physical response. However, their emphasis of these prior studies is spectral combinations and image features, rather than an explicit target-centered partition whose regions provide pooled values for predicting a complete local response vector.

Spatial adaptation has an independent history. Lensen et al. evolve image regions, extract histogram-of-oriented-gradient features, and construct classifiers within a genetic-programming framework[23]. Malinowski and Fritz learn task-dependent pooling regions, while Liu et al. develop adaptive spatial pooling for image classification[24, 25]. These approaches establish precedents for learning where and how image information is aggregated, although their objectives and outputs differ from local spectroscopic regression. Bakurov et al. further demonstrate evolutionary construction of compact image-quality expressions evaluated against continuous human opinion scores[26].

In microscopy, structure–property prediction itself is well established by now. Image-to-spectrum encoder–decoder models connect domain patterns to switching responses, and deep kernel learning connects local STEM images to spectroscopic behavior and guides autonomous measurements[10, 11]. Curiosity-driven sampling provides another route to learning image–response relationships efficiently[9]. The unresolved question addressed here is the construction of an explicit, compact spatial measurement representation. We have not identified a demonstration that jointly evolves target-centered spatial partitions and regional mean, minimum, and maximum pooling for prediction of complete local spectra or hysteresis loops, with descriptor dimensionality as a competing objective. Hence, the central contribution of this study is this constrained descriptor construction and its evaluation for structure-property discovery problems.

## III. Data and regression benchmarks

We focus on two public microscopy datasets: scanning transmission electron microscopy with electron energy-loss spectroscopy (STEM–EELS), and piezoresponse force microscopy with band-excitation piezoresponse spectroscopy (PFM–BEPS). The STEM–EELS and PFM–BEPS datasets were obtained from the public data sets by Roccapriore et al.[11] and Liu et al.[10], respectively. In each case, an image neighborhood centered on a measurement location is paired with the complete local response. STEM–EELS is the principal example for adaptive descriptor construction, while PFM–BEPS provides a complementary test with a different response and spatial organization. Figure 1 introduces the images, measurement locations, and representative measured and predicted responses. Dataset preparation, acquisition metadata, and regression settings are provided in the Supplementary Information and accompanying notebooks.

### III.A. Comparison of regression algorithms

As a first step, we compare Ridge regression, radial-basis-function kernel Ridge regression, Random Forest, a deep convolutional neural network (DCNN), and deep kernel learning (DKL)[27-31]. This benchmark spans linear, kernel, ensemble, and learned-representation approaches. Figure 1 compares representative predictions with the measured responses, while the test-error distributions are retained in Figure S1. The models reproduce the broad spectral and switching behavior to different degrees, with discrepancies depending on both the response and the selected neighborhood. Detailed architectures, optimization settings, data partitions, and quantitative benchmark tables are provided to the Supplementary Information and notebooks. Prediction accuracy is evaluated using the root-mean-square error (RMSE) over measurement centers and response channels. Here we use its normalized form (NRMSE), obtained by dividing RMSE by a reference response variation defined in the Supplementary Methods.

The relative performance depends on the measurement. Random Forest and DCNN provide the best regressor performance in the displayed STEM–EELS sweep, whereas DCNN performs best in the displayed PFM–BEPS comparison. The predicted hysteresis loops capture the broad switching form but retain discrepancies for more asymmetric responses. Ridge and the tested kernel Ridge configuration perform poorly for the PFM–BEPS representation used here. This result motivates examining the descriptors themselves: poor performance with one input representation does not establish that the same regression model will fail with a differently constructed representation..

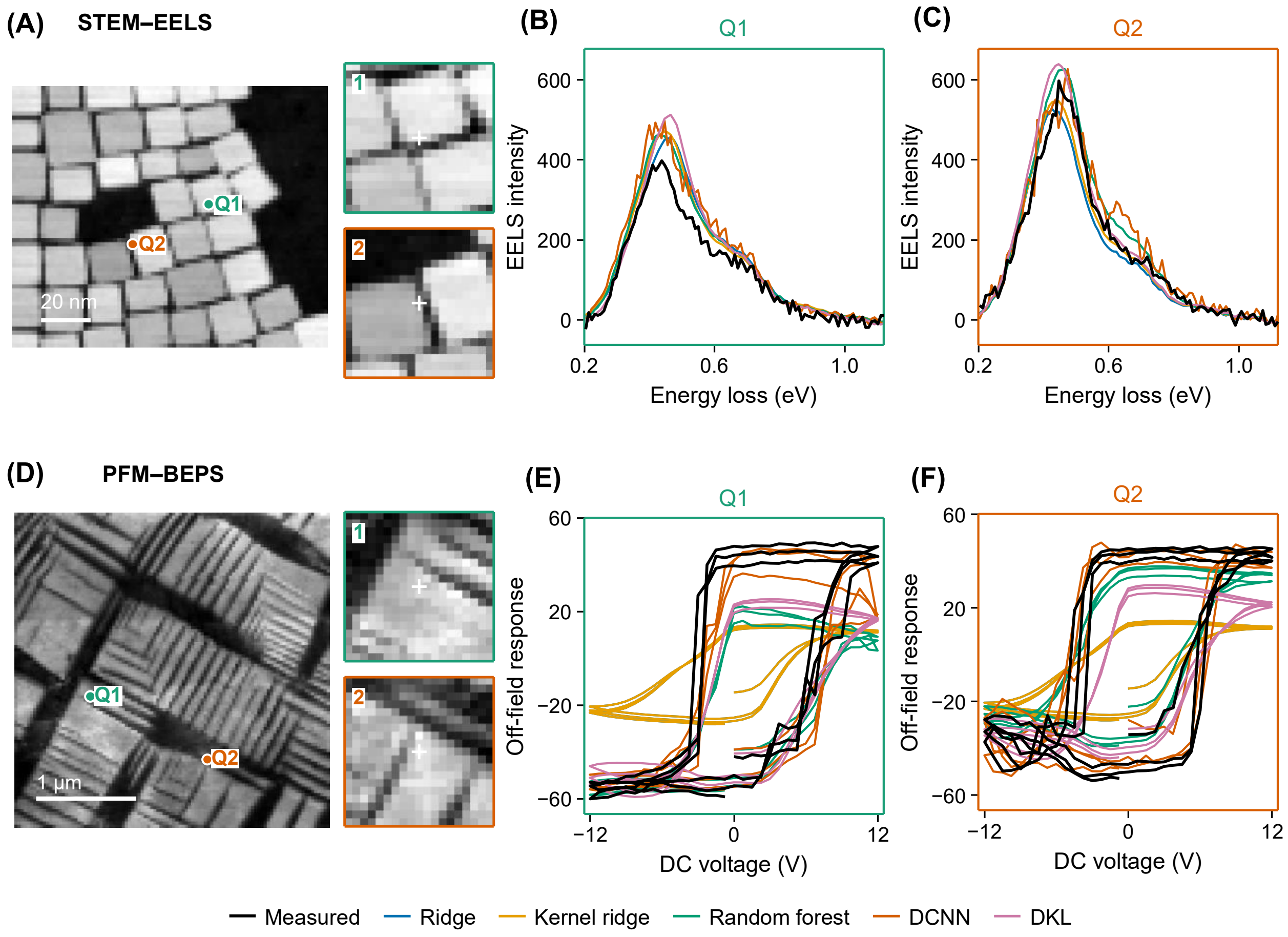

**Figure 1. Image-to-response regression in STEM–EELS and PFM–BEPS. (A, D)** Images with numbered measurement centers and the corresponding extracted neighborhoods. **(B, C)** Measured EELS spectra and predictions from five regressors at centers 1 and 2. **(E, F)** Measured off-field BEPS responses and predictions at the corresponding centers. Black curves denote measurements; model colors are consistent across panels. Scale bars are 20 nm and 1 µm. The test-error distributions are shown in Figure S1.

### III.B. Spatial attribution and its relation to image structure

To explore the origin of the predictive signal, we examine the SHAP attribution maps to find which parts of the image neighborhood contribute to the model's prediction. SHAP maps characterize how the trained predictor assigns contributions within an image neighborhood. Figure 2 focuses on PFM–BEPS and compares a representative input patch, its SHAP importance map for the trained DCNN predictor, and the mean attribution map. Both the individual and average maps emphasize the target neighborhood, while the surrounding patterns vary across locations. This variation motivates examining the distribution of attribution maps rather than interpreting only its average.

Principal component analysis identifies recurring patterns of variation in the aligned SHAP maps. The displayed first component is concentrated near the center, whereas the second contains a broader pattern with contributions of opposite sign. The signs refer to PCA loading directions and do not directly identify positive or negative SHAP contributions without specifying the preprocessing.

Each explained patch has a component score that can be mapped to its measurement center. Spatial score maps therefore connect variation in the attribution patterns to variation in the imaged structure.

Radial importance profiles, explained variance, and canonical correlation analysis (CCA) between image-patch and SHAP representations are presented in Figure S2. CCA tests whether combinations of image features covary with combinations of attribution features; it does not independently establish a causal structure–property relationship. The cross-validation results vary substantially across folds, so the displayed correlations alone do not support a robust general association. Overall, the SHAP – PCA analysis suggests that primary origin of signal is local, with the additional significant components associated with the proximity to the domain walls as can be expected from the physics of ferroelectric switching.

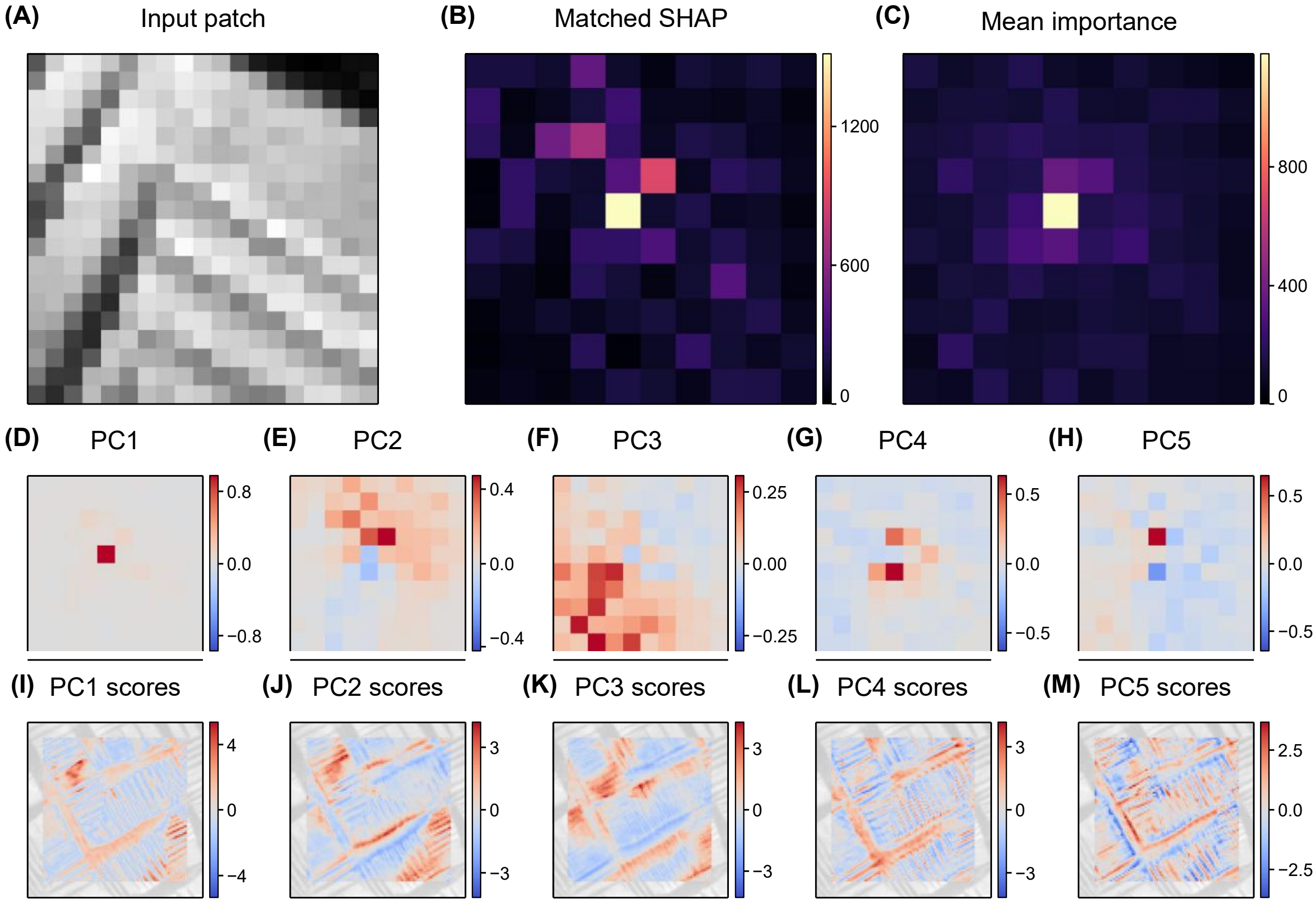

**Figure 2.** Spatial attribution and PCA for PFM–BEPS. **(A)** Representative input patch for the DCNN. **(B)** Matched SHAP map and **(C)** mean importance map. **(D-H)** Loading maps of Principal Components 1 to 5, accounting for 37.4%, 6.82 %, 5.4%, 3.9% and 3.0% of the reported variance. **(I-M)** Corresponding standardized scores at all the image-patch locations. Loading maps and scores for components 6-20 are represented in Figure S2. Radial importance, explained variance, and CCA are presented in Figure S3.

### III.C. Spatial extent and uniform compression

We next separate the extent of the neighborhood from the resolution supplied to the regressor (Figure 3). A square patch of side N is compressed by block-mean pooling with factor m, producing k × k values, where the supplied implementation uses k = ceil(N/m). Increasing N changes the surrounding structure available to the model. Increasing m reduces the detail retained within that neighborhood.

The grids show that useful resolution depends on the dataset and the regressor. In STEM–EELS, several compressed neighborhoods retain substantial predictive information, whereas increasing the number of input values does not consistently improve performance. PFM–BEPS exhibits a different balance between neighborhood size and compression (Figure S4). The retained feature count therefore cannot be selected independently of spatial extent, model class, and the measured response. This uniform compression is a useful baseline because it separates spatial context from retained resolution with few parameters. Compression can preserve useful predictive information while reducing dimensionality, but its benefit must be assessed against the corresponding uncompressed neighborhood under matched evaluation conditions. The observed variation motivates a representation in which different regions retain different levels of detail.

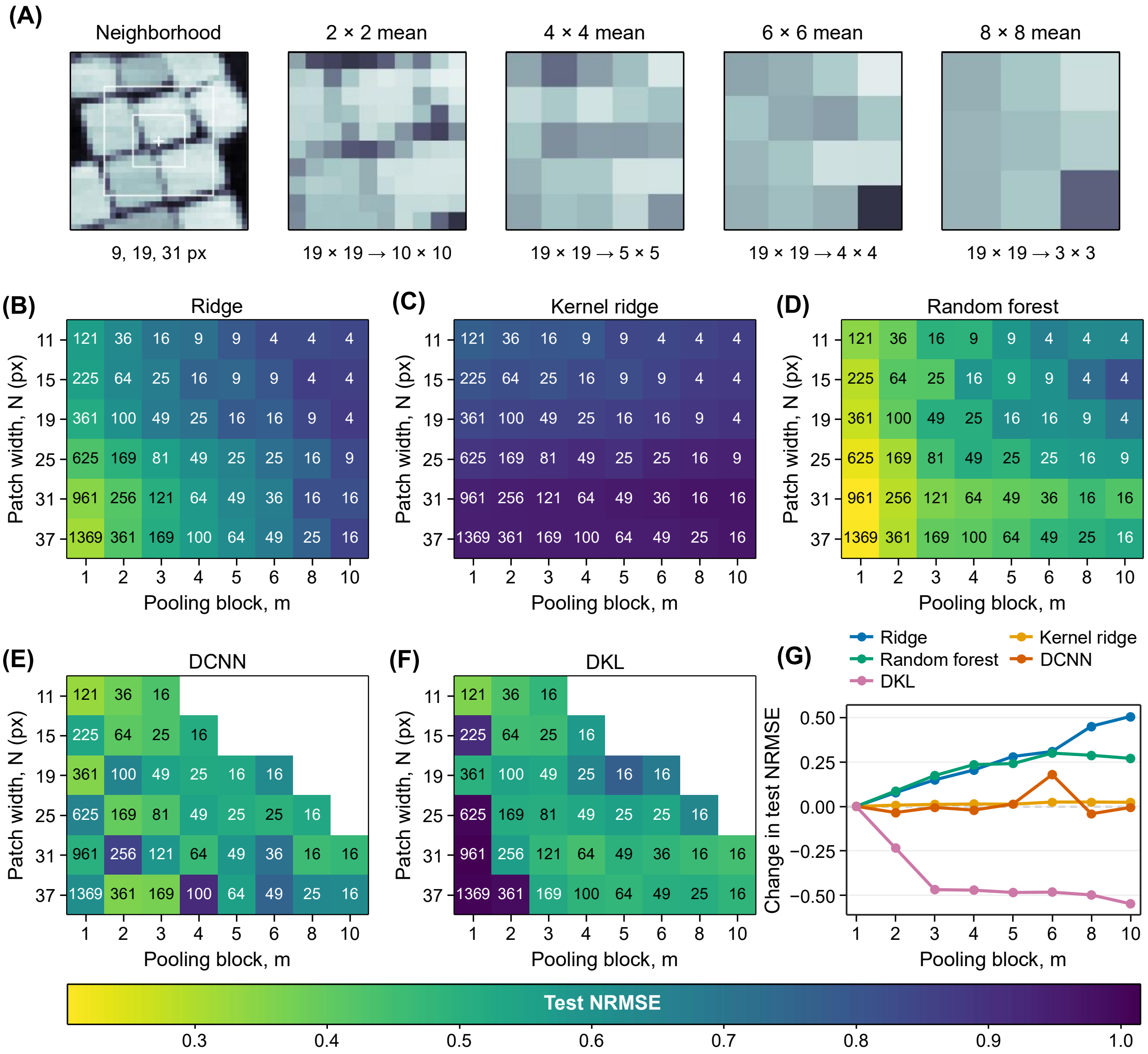

**Figure 3. Spatial extent and uniform compression in STEM–EELS. (A)** Neighborhood selection and mean pooling of a 19 × 19 patch using 2 × 2 and 4 × 4 blocks. Spaces are reserved for compression factors 6 and 8. **(B–F)** Test NRMSE for Ridge, kernel Ridge, Random Forest, DCNN, and DKL as a function of patch width and compression block size. Numbers inside the cells indicate retained scalar features. Blank cells indicate unavailable configurations in the source figure. BEPS grids and selected error curves are shown in Figure S4; performance at compression factors above 4 is not yet supplied.

### III.D. Prediction error and descriptor dimensionality

Figure 4 represents the grid results in terms of prediction error and the number of scalar inputs, directly visualizing the Pareto front between the two. A descriptor is nondominated if no evaluated alternative has both equal or lower error and equal or lower dimensionality, with a strict improvement in at least one quantity. The resulting front identifies useful compromises, while the full point cloud preserves information about configurations that add dimensions without improving prediction. High-complexity configurations on an upward error branch are dominated and remain outside the strict Pareto front. Figures 4 and S4 summarize reported test errors across evaluated configurations. These comparisons are descriptive rather than validation-based selection followed by an independent test assessment.

This representation allows models to be compared at a common feature budget. Figure 4 places the STEM–EELS and PFM–BEPS fronts side by side, while the corresponding fixed-budget summaries appear in Figure S5. The main comparison separates changes in predictor performance from changes in the number of available inputs. Descriptor dimensionality remains distinct from model size, feature-extraction cost, and inference time: fewer inputs do not by themselves establish a faster workflow.

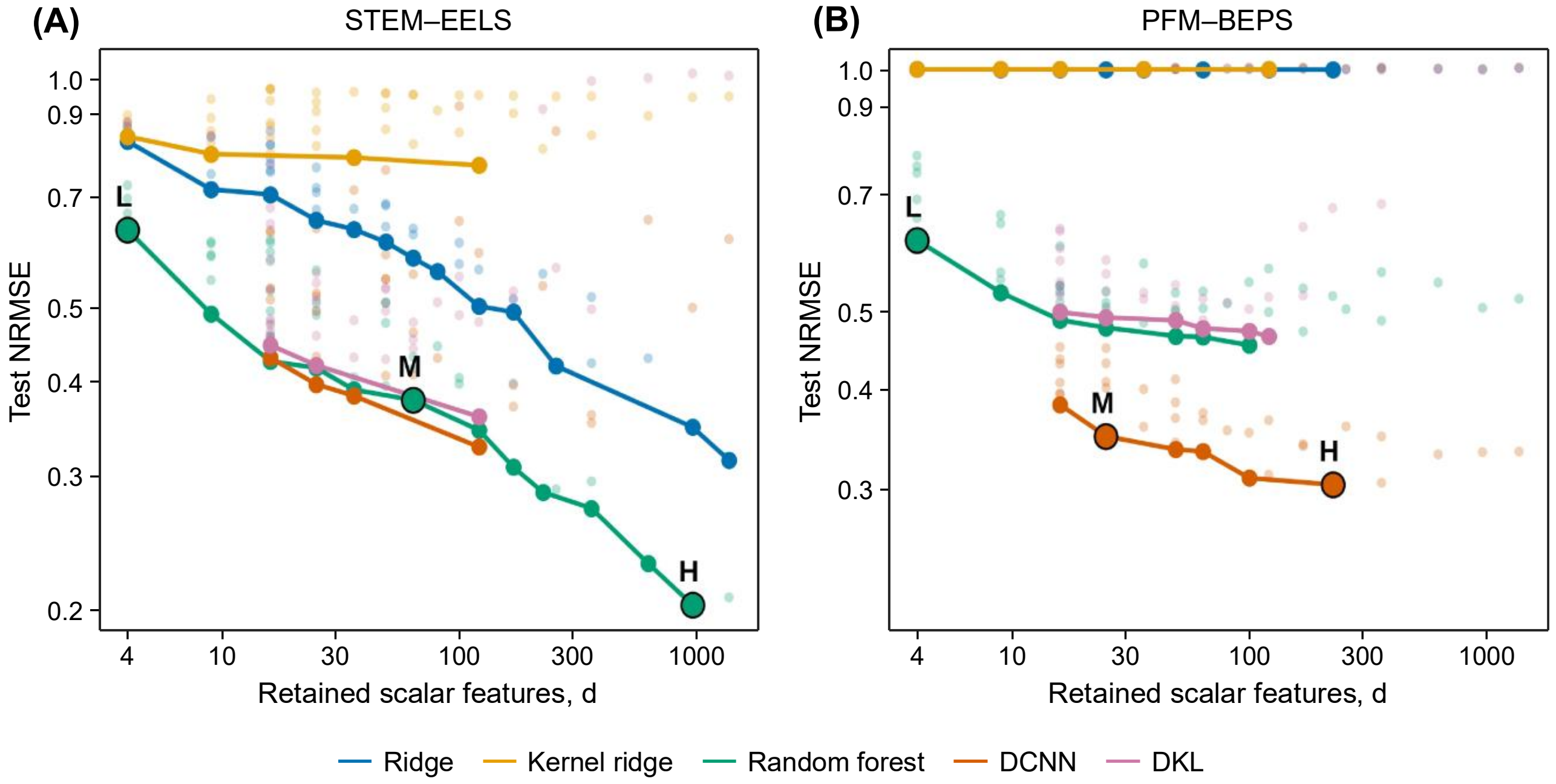

**Figure 4. Pareto front between descriptor complexity for prediction accuracy.** Reported test NRMSE versus retained descriptor values for **(A)** STEM–EELS and **(B)** PFM–BEPS. Faint points show all evaluated configurations, revealing how prediction performance varies across the descriptor search. Colored lines trace each regressor's nondominated front: configurations offering lower error than any evaluated alternative with fewer descriptor values. L, M, and H highlight representative low-, intermediate-, and high-dimensional solutions. Fixed-budget summaries are shown in Figure S5.

## IV. Genetic construction of spatial descriptors

Here we use genetic algorithm to optimize the measurement representation supplied to a separately fitted regressor. A candidate, or chromosome, specifies a partition of a target-centered image neighborhood and a pooling operation for each region. Applying the same chromosome to every patch yields a descriptor matrix; fitting the regressor on this matrix and evaluating the complete predicted responses yields the candidate's predictive error. Evolution changes the partition and regional statistics, while the regression model is refitted for each candidate. The two objectives are low validation error and a small number of retained scalar values. Thus, the search produces a family of accuracy–dimensionality compromises rather than imposing a single compression factor.

**IV.A. Descriptor encoding and evolutionary procedure**

Uniform pooling imposes the same resolution throughout a patch. The genetic representation instead specifies a set of spatial regions and an operation applied to each region (Figure 5). Mean, minimum, and maximum pooling each contribute one descriptor value; discarded regions contribute none. Two adjacent maximum-pooled regions are two features, whereas their merged union summarized by one maximum is one feature. The black boundaries in the operation masks therefore encode feature identity, while the colors encode the statistics within the region.

In the Cartesian construction, internal nodes of a partition tree represent subdivisions and terminal nodes identify regional measurements. Refinement replaces a terminal region with smaller children, increasing the available local resolution. Coarsening replaces a set of descendants with their parent region and a single pooling operation. Operator mutation changes a terminal region between mean, minimum, maximum, and discard without changing its support; reactivating a discarded region restores a scalar feature. Coarsening is restricted by the hierarchy so that adjacent regions cannot necessarily be merged unless they belong to a collapsible subtree. Crossover exchanges compatible subtrees between two parent chromosomes, transferring a spatial arrangement together with its regional operations while preserving the partition constraints.

For polar supports, the neighborhood is discretized into four radial bands and eight angular bins. A chromosome groups the elementary bins into connected regions and assigns mean, minimum, maximum, or discard to each region. Merging combines neighboring regions; connected splitting restores finer support; and operator mutation changes the regional statistic or discards a region. Angular adjacency wraps across the angular boundary. Region-transfer crossover inserts a donor region into a receiving partition, removes its cells from the receiving regions, and separates any disconnected remainders into connected components. These operations preserve a complete, valid partition, while discarded regions contribute no descriptor values (Figure 6).

Evaluation is nested inside evolution. Each candidate is applied consistently to all patches, the downstream regressor is fitted using the fitting subset, and validation error is computed over the complete response vector. The feature-count objective is the number of retained regional values, not the number of pixels covered. A region summarized by one extremum counts once regardless of its area; a discarded region contributes no value. Selection uses the two objectives to retain useful alternatives across the error–dimensionality front. Selected descriptors are subsequently refitted using the combined fitting and validation subsets before assessment at reserved test locations. Test responses do not determine genetic operations or survival. Test responses are not used for candidate fitness, genetic operations, survival, or selection of the displayed descriptors.

The figure-producing searches begin with simple global summaries, finely resolved representations, and random partitions. After candidate evaluation, the lowest-validation-error individual is retained, and additional survivors are selected through tournaments using Pareto rank and crowding distance. Pareto rank favors candidates not dominated in error and feature count; crowding distance favors underrepresented parts of a front. Surviving parents generate offspring by mutation and crossover, while random immigrants maintain exploration. Evaluation and selection then repeat, and the archive retains evaluated candidates so that a useful nondominated descriptor is not lost merely because it leaves the active population. Repeated searches characterize optimization repeatability on the fixed data partition; their settings and stopping criteria are specified in Supplementary Methods.

### IV.B. Cartesian descriptors

Figure 5 shows the resulting Cartesian search. The validation front improves rapidly as a small number of regional summaries is introduced, followed by more gradual gains at larger dimensionality. The simplest illustrated descriptor averages the entire patch. The intermediate descriptor combines large maximum-pooled regions with smaller regions near the center, while the higher-dimensional example retains finer subdivisions. Over much of the displayed feature-count range, the adaptive front lies below the uniform-pooling reference. This comparison indicates that the placement of spatial resolution and the choice of regional statistic can matter beyond the total number of retained values.

The validation front guides representation selection, whereas test errors characterize selected descriptors at reserved locations. Most of the improvements occur between the simplest and intermediate examples, with a smaller additional gain from the more detailed partition. The bend in the front therefore identifies a useful compact representation for this dataset. It is an operating point within the explored family, rather than a universal optimum across specimens or regression models.

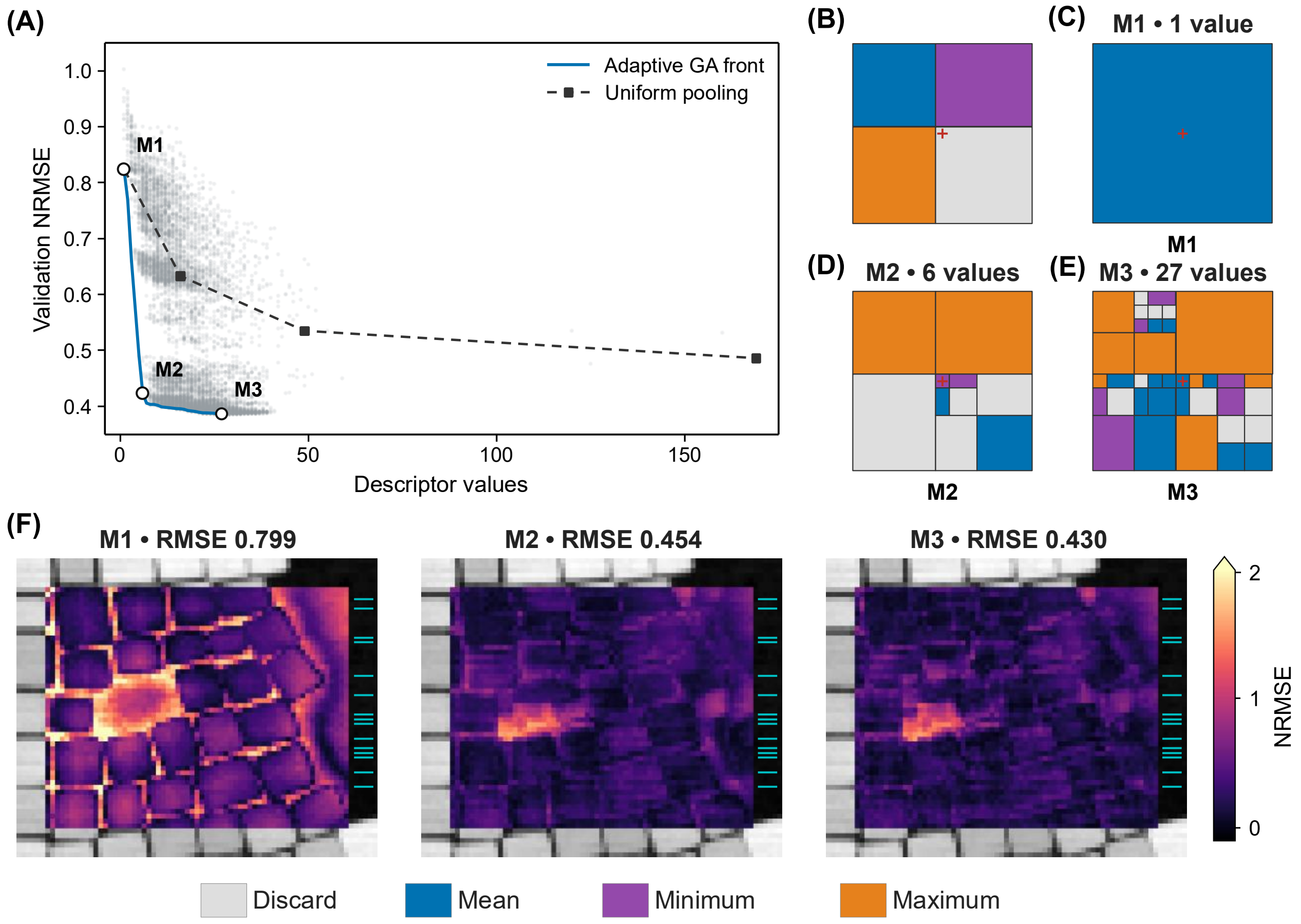

**Figure 5. Genetic construction of Cartesian descriptors. (A)** Validation-error front for 11,200 distinct recipes across three seeds, compared with uniform pooling. **(B)** Cartesian subdivision and pooling operations. **(C–E)** Selected masks containing 1, 6, and 27 values, with test NRMSE values of 0.799, 0.454, and 0.430. M1, M2, and M3 on the front identify these three candidates. Colors encode discard, mean, minimum, and maximum operations. **(F)** Full-image error maps for the selected masks. Error distributions and held-out diagnostics are in Figure S6.

Figure S6 provides held-out error distributions, cumulative distributions, channel-wise errors, measured-versus-predicted intensity plots, and example spectra. These diagnostics connect aggregate errors to individual responses. Figure 5F instead maps errors at all locations, including training and test, with held-out rows explicitly marked. Here, the errors at training locations are fitted-data diagnostics whereas a prediction at held-out centers is classical test set. Note that error distribution is uniform across the maps without clear variations between training and test rows.

### IV.C. Polar descriptors and alternative supports

Polar coordinates organize a neighborhood by distance and angle from the measurement location (Figure 6). The circular masks therefore provide an interpretable description of how radial and directional information is retained. Circular support excludes the corners of the corresponding square patch, so coordinate system and available image support are separate experimental choices.

Figure 6 shows the polar search and representative descriptors along its validation front. The simplest illustrated candidate retains a small central sector and discards most of the disk. The intermediate candidate uses broader radial and angular summaries, while the more detailed candidate partitions the disk more finely. As in the Cartesian example, the largest improvement occurs on moving beyond the simplest representation. The masks make the spatial consequences of retaining additional descriptor values directly visible.

The displayed polar encoding uses four radial bands and eight angular bins with discard, mean, minimum, and maximum operations; its unwrapped representations are shown in Figure S9. Angularly resolved descriptors retain directional information, so circular support alone does not make a descriptor rotationally invariant. Although both displayed searches use the same regression family, their selected errors do not by themselves isolate the effect of geometry. Such a comparison also requires matched spatial support, data partitions, preprocessing, and search budgets.

Three-seed fronts and convergence traces for both geometries are shown in Figure S9. Their variation describes optimization repeatability on fixed data, rather than uncertainty across independent images. Figure S7 provides held-out error distributions, cumulative distributions, channel-wise errors, measured-versus-predicted intensity plots, and example spectra. Figure 6F maps errors at all 3,150 eligible centers, including training and test locations, with held-out rows explicitly marked. Together, these diagnostics relate aggregate prediction errors to their spectral and spatial variation.

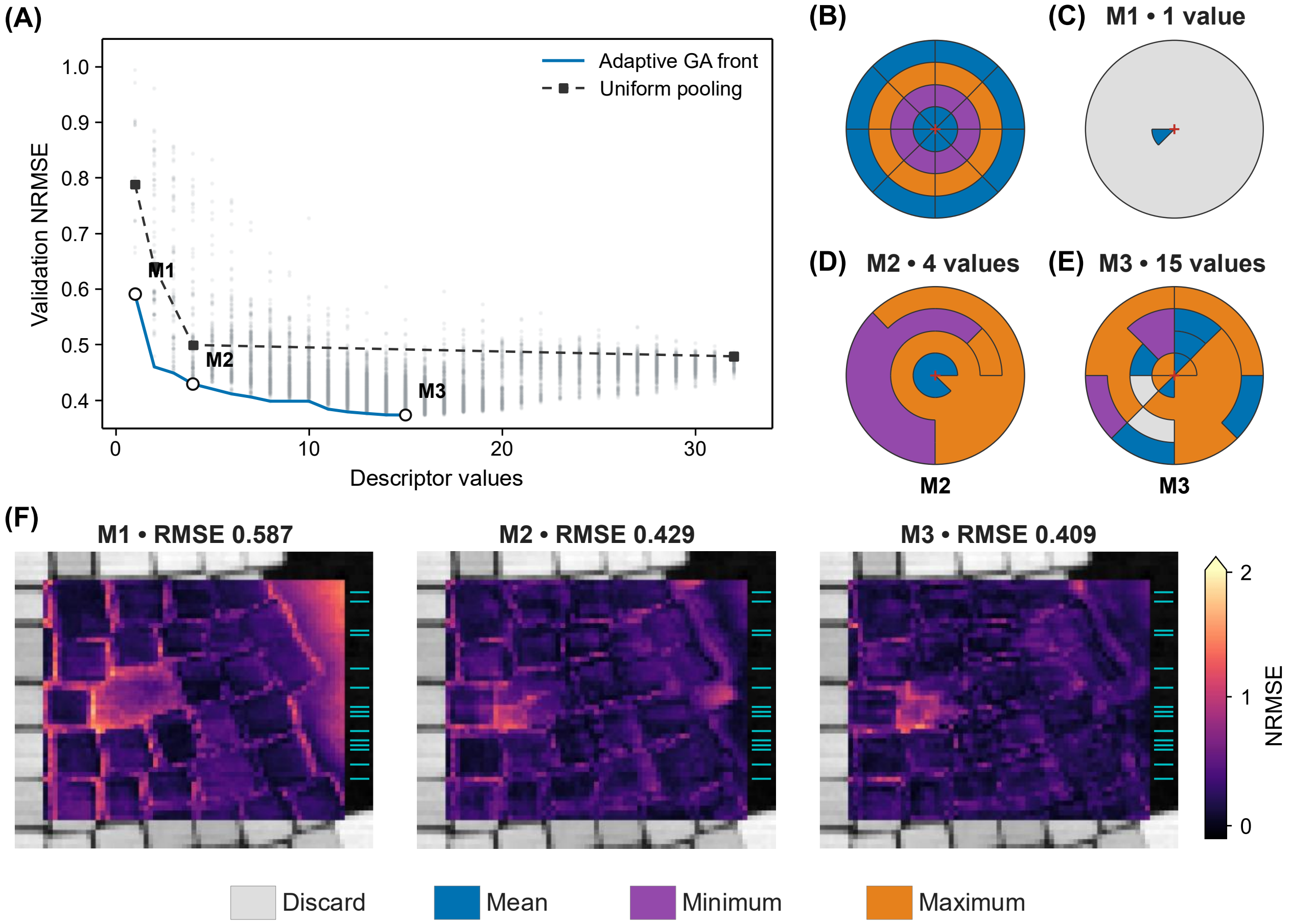

**Figure 6. Genetic construction of polar descriptors. (A)** Validation-error front for 13,724 distinct recipes across three seeds, compared with uniform pooling.. **(B)** Four radial bands and eight angular bins **(C–E)** Selected masks containing 1, 4, and 15 values, with test NRMSE values of 0.587, 0.429, and 0.409. M1, M2, and M3 on the front identify these candidates. Colors have the same meaning as in Figure 5. **(F)** Full-image error maps for the selected masks. held-out diagnostics, optimization repeatability, and unwrapped masks appear in Figures S6–S8.

## V. Alternative descriptor strategies

The partition-based construction is one member of a broader family of spatial descriptors. We also evaluated connected regions without a tree constraint, overlapping region dictionaries, and alternative region and footprint geometries. Figure 7 illustrates selected constructions using unchanged square pixels. Multiscale annuli and additional regional statistics that capture variation or contrast remain prospective extensions. A separate archived study evaluated discrete symmetry averaging, but those results are not included in the present figures. These strategies change the information retained by the representation and should be compared using the same response, data partition, regressor, and complexity measure.

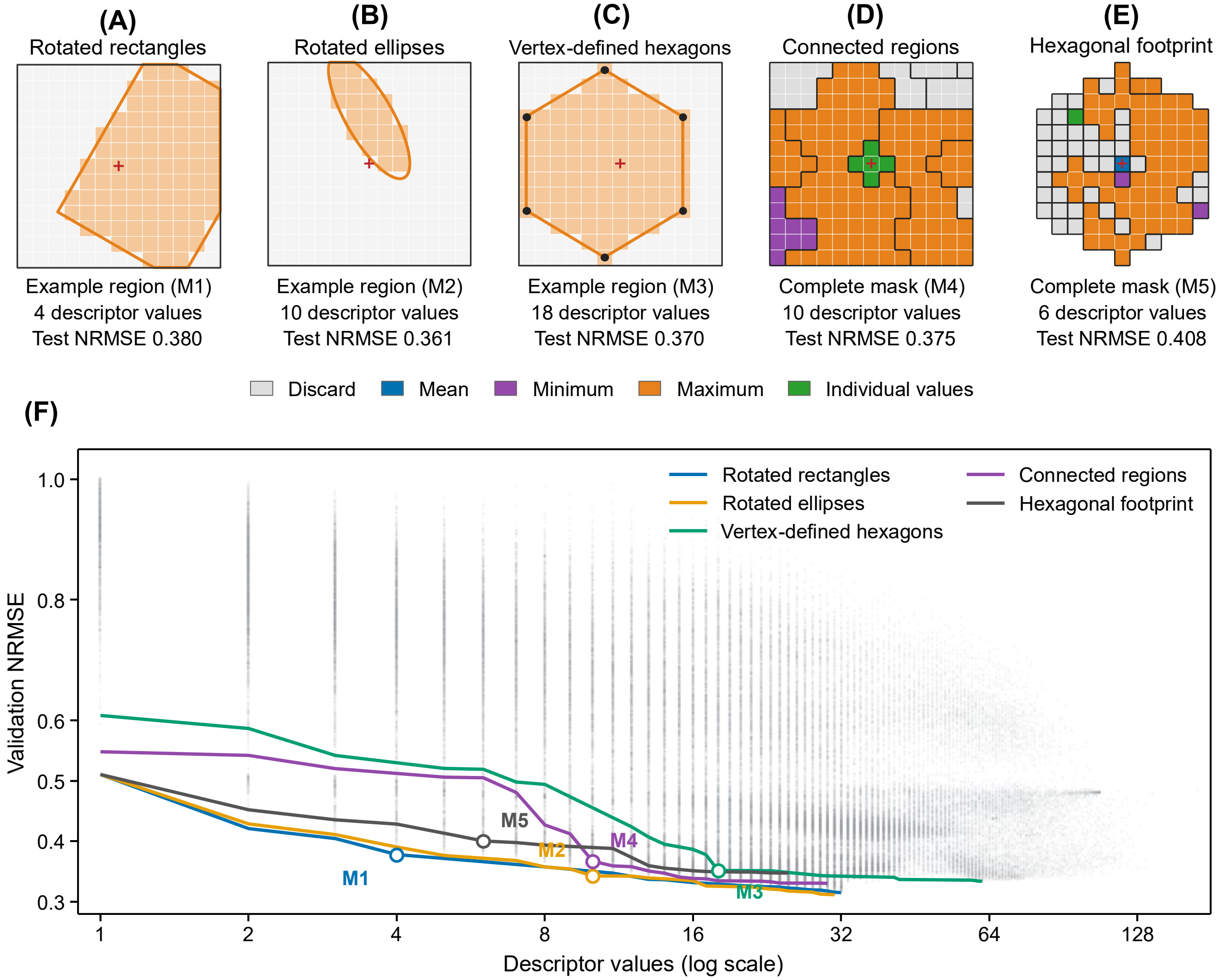

**Figure 7. Alternative spatial descriptors and their prediction errors. (A–C)** Example selected rectangular, elliptical, and vertex-defined hexagonal regions. **(D, E)** Complete optimized masks using connected regions and a hexagonal footprint on unchanged square pixels. Colors encode discard, mean, minimum, maximum, or retention of individual values. M1–M5 identify the complete descriptors; their sizes and test NRMSE values are shown below. **(F)** Validation Pareto fronts with evaluated recipes shown faintly and M1–M5 marked.

Beyond the Cartesian and polar constructions, we explored rotated rectangular and elliptical regions, vertex-defined hexagons, connected regions, and hexagonal footprints defined on the original square-pixel grid (Figure 7). These constructions vary the spatial information retained without changing the measured pixel values. Selected descriptors containing 4–18 values achieved test NRMSE values of 0.361–0.408, compared with 0.496 for Ridge regression using all 169 patch pixels. Error maps show how full-spectrum prediction errors vary across the held-out locations. These exploratory comparisons demonstrate the flexibility of automated descriptor construction; differences in descriptor length, search space, and optimization budget prevent attributing performance differences to region shape alone.

## VI. Discussion

We propose that image representation can be optimized as part of the regression problem. Spatial extent determines what surrounding structure is available, while regional partitioning and pooling determine how this structure enters the predictor. The supplied examples support both model-dependent gains and substantial dimensionality reduction. They do not support a universal preference for the largest patch, the smallest descriptor, or a particular geometry.

The distinction from symbolic regression lies in the evolving object. PySR searches mathematical expressions, and SISSO selects a small set of generated symbolic features. Here, the search selects spatial supports and the measurements made over those supports, while the downstream regressor is fitted separately. Minimum and maximum pooling introduce nonlinear transformations of the image even when the regressor is linear in its input features. This explains how changing the descriptor can improve Ridge without increasing the complexity of the regression model itself.

The masks are interpretable because every feature has explicit support and statistics. Their physical meaning nevertheless depends on the observational process and the correlations in the dataset. Several image regions can contain interchangeable predictive information, and different masks can yield comparable errors. A stable radial range or recurring morphological feature may therefore be more informative than identical boundaries across stochastic searches. Spatial attribution and optimized masks address related but distinct questions: SHAP describes a fitted predictor, whereas the genetic search changes the representation that the predictor receives.

## VII. Conclusions

We develop a framework for constructing explicit spatial descriptors for local structure–property prediction. Regression benchmarks and SHAP maps expose the dependence on both the predictor and its representation. Uniform compression separates spatial context from retained resolution, and genetic search produces nonuniform regional measurements with a controllable number of inputs. In the EELS examples, Cartesian and polar descriptors with a small number of values capture much of the improvement available along their respective validation fronts. The approach provides a way to test which spatial summaries preserve the information needed for a specified response, while keeping the retained regions accessible to inspection and physical interpretation.

With larger collections of paired images and local responses, the same framework can be extended across specimens, material classes, and spatial resolutions. Expressing region sizes in physical units and testing descriptors across independently acquired images would distinguish specimen-specific patterns from reproducible spatial ranges. For related measurement problems, joint optimization over multiple datasets may yield reusable descriptor families that retain the relevant context while adapting to pixel size and instrumental resolution. Recurrent supports could then identify characteristic length scales of predictive structure–property coupling, provided that instrumental sensitivity and correlated microstructure are separated from the underlying physical interactions. Such broadly applicable descriptors are a prospect enabled by sufficiently diverse structure–property data, rather than an assumption of the present within-dataset study.

A further extension is to molecular structures and solids, where the image partition would be replaced by a chemically and geometrically admissible decomposition. For molecules, candidate operations could select atomic environments, merge adjacent atoms into connected fragments, split fragments, and pool atom or bond attributes using means, extrema, or other defined summaries. Crossover could exchange descriptor fragments while preserving connectivity and a consistent feature ordering; because the representation rather than the molecule is evolved, these operations need not alter the chemical structure. For solids, analogous operations could group sites by species or coordination environment, merge or split coordination shells, change neighborhood cutoffs, and aggregate bond lengths, angles, or local structural invariants. Translation, rotation, permutation symmetry, and periodicity would constrain the allowed operations. Establishing these rules would make it possible to search for interpretable molecular and solid-state representations using the same predictive-error–complexity principle.

**Acknowledgements**
This research is supported (A.R and S.V.K) by the National Science Foundation Materials Research Science and Engineering Center program through the UT Knoxville Center for Advanced Materials and Manufacturing (DMR-2309083).

**Data and code availability**
The datasets used in this study originate from the public data associated with Refs. [10, 11]. The input data, saved results, and notebooks for reproducing the analyses are available at https://github.com/adityaraghavan98/Interpretable-Spatial-Descriptors.

## Supplementary methods

The STEM-EELS image and spectrum are the dataset from Ref. 7, with 1.6045 nm pixel spacing (62 × 75 px, ~99 × 120 nm field of view) and an energy axis spanning 0.200–1.132 eV at 0.000623 eV steps (1,496 channels before the 15-channel binning described below). The PFM-BEPS amplitude image and off-field loops are derived from per-pixel SHO-fit maps of the $PbTiO_3/KTaO_3/SrRuO_3$ dataset from Ref. 8 (100 × 100 px, 3 × 3 μm scan, 30 nm pixel spacing), with piezoresponse amplitude reported in millivolts and a DC bias sweep spanning −12 to +12 V across 192 loop points. Image and response channels share the same pixel grid for both datasets, so no separate spatial registration step is applied.

The input at each measurement center is an N × N image neighborhood paired with its full local response. STEM–EELS uses a 62 × 75-pixel image and 1,496-channel spectra. Non-overlapping averaging of groups of 15 channels gives 99 response channels; the trailing 11 channels are discarded. PFM–BEPS uses a 100 × 100-pixel amplitude image and 192-point off-field hysteresis loops without response-axis binning. Images and response cubes receive no normalization, denoising, or background subtraction before patch extraction.

Uniform compression averages non-overlapping m × m blocks after edge-padding each patch to a multiple of m, giving k × k inputs with $k = \lceil N/m \rceil$. Eligible centers exclude a border of $\lfloor N/2 \rfloor$ pixels, so eligible centers differ across patch sizes. Classical regressors use N ∈ {11, 15, 19, 25, 31, 37} and m ∈ {1, 2, 3, 4, 5, 6, 8, 10}: 48 configurations in total. Convolutional and deep kernel regressors use the union of this grid and N ∈ {5, 9, 13, 19, 25, 31} × m ∈ {2, 4}, restricted to k ≥ 4, for 39 configurations total. For PFM–BEPS, candidate N values are additionally restricted below the shorter image dimension, which has no effect at 100 × 100. Their embedding widths remain fixed at 20 and 16, respectively, chosen near previously identified best settings rather than searched alongside N and m.

At each configuration, a grouped random split assigns 30% of eligible centers to test using seed 0. Grouping by image-center row keeps centers from the same row together, and all regressors use the same partition. NumPy and PyTorch random generators are initialized with seed 0. The uniform-compression sweep has no independent validation subset: its reported best configurations minimize error on the test subset itself.

Classical regressors receive flattened, unscaled compressed patches. Response channels are centered by training-subset means, restored before evaluation. Ridge uses regularization strength 1.0; kernel ridge uses a radial-basis-function kernel with coefficient $\gamma$ equal to the inverse feature count; random forest uses 50 trees and at least three samples per leaf. Neural inputs are standardized with a single training-subset image mean and standard deviation, while response channels are standardized separately. All predictions are returned to original response units before evaluation.

The convolutional regressor predicts the full response vector and is trained for 50 epochs with batch size 256 and weight averaging. This budget was reduced from 250 after training and test loss curves showed minimum test loss near epochs 15–20. Test data were supplied for monitoring during training. Every configuration receives the fixed budget without early stopping.

Deep-kernel regression combines a convolutional encoder, with two rounds of local filtering and spatial down sampling, and a sparse variational Gaussian process. PCA fitted to standardized training responses retains five components; their scores are standardized again using training statistics. Each component has a separate, independently initialized and trained encoder–Gaussian-process model. Training maximizes a variational evidence bound for 120 full-batch gradient iterations without early stopping. Predictions are reconstructed by reversing score standardization, PCA, and response standardization.

Root-mean-square error (RMSE) is calculated by squaring the differences between predicted and measured responses, averaging over all evaluated measurement centers and response channels, and taking the square root. Normalized root-mean-square error (NRMSE) divides RMSE by a reference response variation. For test assessment, this reference is the root-mean-square deviation of measured test responses from their channel-wise test-subset means. For GA validation, it is the root-mean-square deviation of measured validation responses from the fitting-subset channel means.

Per-location NRMSE is the RMSE across response channels at one measurement center divided by the same subset-level reference. Aggregate NRMSE is the root mean square of these per-location values, rather than their arithmetic mean. Channel-wise RMSE remains in the original response units. The coefficient of determination is weighted by response-channel variance. The minimum test-NRMSE configuration for each model and dataset is used in the best-setting benchmark and SHAP analysis; these minima are descriptive comparisons rather than independent validation-based selections.

Attribution analyses use all patches used in the regressors: 2,508 for STEM–EELS and 6,724 for PFM–BEPS. STEM–EELS explains the random forest using 19 × 19 patches pooled by a factor of four, giving 25 features and 99 response channels. PFM–BEPS explains the DCNN using 19 × 19 patches pooled by a factor of two, giving 100 features and 192 response channels. STEM–EELS uses Tree-SHAP without an explicitly supplied background array. PFM–BEPS uses permutation SHAP with 60 training-background patches and a maximum evaluation budget of 500 per explained patch. Predictions are returned to original response units before attribution.

Each importance descriptor sums absolute SHAP contributions across response channels. Mean maps average these descriptors across all eligible patches. For radial importance, radial profiles first normalize each patch descriptor to 100% total importance, then average feature percentages within radial bins and across patches. Bin widths equal the pooling factor. Radii use pooled-feature centers accounting for edge padding, with native pixel spacings of 1.6045 nm for STEM–EELS and 30 nm for PFM–BEPS. Shaded bands show the across-patch interquartile range.

Attribution PCA uses all centered, unstandardized importance descriptors. Figure 2 displays the first five PFM–BEPS loading maps and spatial scores, standardized using full-dataset score means and standard deviations. Image–SHAP CCA uses separate three-component PCAs of pooled image intensities and importance descriptors, followed by two-component CCA with internal scaling and at most 5,000 iterations. Figure S3 displays the first canonical pair fitted to the full dataset. Five-fold cross-validation groups centers by image row and refits both PCAs and CCA within each fitting fold. Reported cross-validation correlations are held-out fold means ± sample standard deviations. Overlapping neighborhoods remain possible across folds, and these descriptive associations do not establish causality or independent-specimen generalization.

The GA comparisons use the same STEM–EELS data and binning, with 13 × 13 patches paired with their center spectra, without spatial averaging of spectra. The 3,150 eligible centers are split by image-center row: the outer split uses test fraction 0.3 and seed 0, giving 2,205 training and 945 test centers. An inner split of the training subset uses validation fraction 0.2 and seed 1, yielding 1,764 fitting and 441 validation centers. Subsets remain fixed across descriptor families and search seeds. Center rows are disjointed, but patches can overlap across subsets; no spatial exclusion buffer is imposed.

The encodings and evolutionary operations follow Section IV.A. Cartesian regions use original pixels without padding or interpolation. Polar descriptors use a radius-six-pixel disk with four radial bands and eight angular bins, each sampled at 3 × 3 interior positions by bilinear interpolation. Regional means combine bin means with annular-area weights proportional to 1, 3, 5, and 7; extrema use sampled intensities. Neither augmentation nor symmetry averaging is applied.

Candidates are evaluated using multi-output Ridge with regularization strength 10.0. Descriptor columns use fitting-subset means and standard deviations; scales below $10^{-12}$ are replaced by one. Responses are centered by fitting-subset channel means without variance scaling or channel-specific weighting. The objectives are pooled full-spectrum validation RMSE and retained descriptor count. After descriptor selection, preprocessing and regression are refitted on all 2,205 training centers.

Cartesian and polar searches use 25 candidates, five survivors, and seeds 0, 1, and 2. Initial populations include global summaries, finely partitioned descriptors, and random candidates. The lowest-validation-error candidate survives, with descriptor count breaking ties; four additional survivors are chosen through three-candidate tournaments favoring lower Pareto rank and then larger crowding distance. Each offspring attempt introduces a random immigrant with probability 0.15. Otherwise, two survivors are sampled with replacement, crossover is attempted with probability 0.8, and mutation follows. Polar mutation probabilities are 0.35 for merging, 0.30 for splitting, 0.25 for pooling changes, and 0.10 for discard. Candidates must retain at least one value. Within-generation duplicates are rejected, repeated chromosomes reuse cached evaluations, and all evaluated recipes are archived.

Searches run for at least 60 and at most 240 generations, stopping after 60 generations without a normalized hypervolume improvement greater than $10^{-5}$ since the last qualifying improvement. Hypervolume uses a reference of 170 values and 1.5 times the validation RMSE of the fitting-mean predictor, normalized by the product of these coordinates. All six runs reach 240 generations. The archives contain 11,200 distinct Cartesian and 13,724 distinct polar chromosomes across three seeds. Identical chromosomes count once; different specifications remain distinct even when their descriptor values coincide.

Fronts retain nondominated validation-error–descriptor-count combinations. Figures 5–6 show the fewest-value and lowest-error candidates and a compromise selected by minimum Euclidean distance to the ideal corner after scaling each front axis to [0, 1]. Uniform references use mean, minimum, and maximum pooling at Cartesian block widths 1, 2, 4, 7, and 13, with edge-padding where needed, and polar partitions of 1 × 1, 2 × 1, 4 × 1, 2 × 4, and 4 × 8 radial–angular bins.

Figure 7 searches use the same data, subsets, preprocessing, and Ridge settings, with constructions described in Section V. Rectangular and elliptical recipes retain at most 32 values. Connected-region descriptors use 39 base cells covering the original 169 pixels; the hexagonal footprint contains 107 whole square pixels. Every search uses 25 candidates, five survivors, and seeds 0–2. Rectangular and elliptical searches run for at least 60 and at most 240 generations, stopping after 60 generations without qualifying improvement. Vertex-defined hexagon, connected-region, and footprint searches run for at least 100 and at most 400 generations, stopping after 100 generations without a qualifying improvement. Rectangular and elliptical searches use normalized hypervolume with a 33-value reference. Other extensions use unnormalized hypervolume with references of 170 values, or 108 for the footprint. All use the same reference error of 1.5 times the validation RMSE of the fitting-mean predictor and an improvement threshold of $10^{-5}$. All reported runs reach their generation limits. The connected-region search starts with the best preceding fixed-cell mask for each seed and retains compatible candidates from that earlier archive, adding cumulative search effort beyond the other extensions. Example spectra are chosen by spatial position rather than prediction error. Figure 7F reports validation fronts, while errors accompanying selected masks are test NRMSE after refitting.

## Supplementary figures

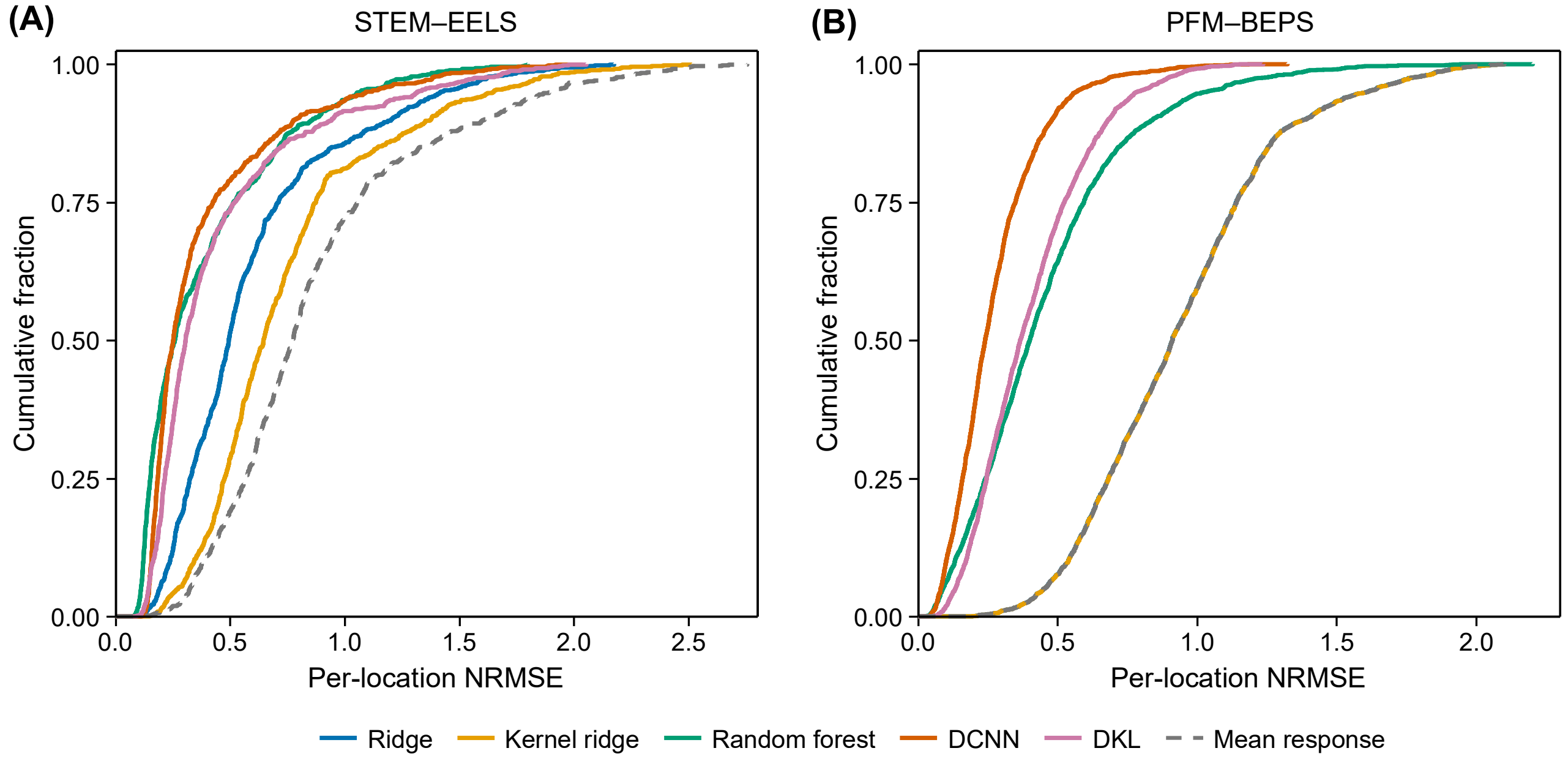

**Figure S1. Cumulative test-error distributions** for (A) STEM–EELS and (B) PFM–BEPS. Model colors follow Figure 1; the dashed gray curve is the mean-response baseline. Curves show empirical cumulative distributions of per-location full-response NRMSE at common test centers within each dataset. Predictions use N = 19, m = 4 for STEM–EELS and N = 19, m = 2 for PFM–BEPS.

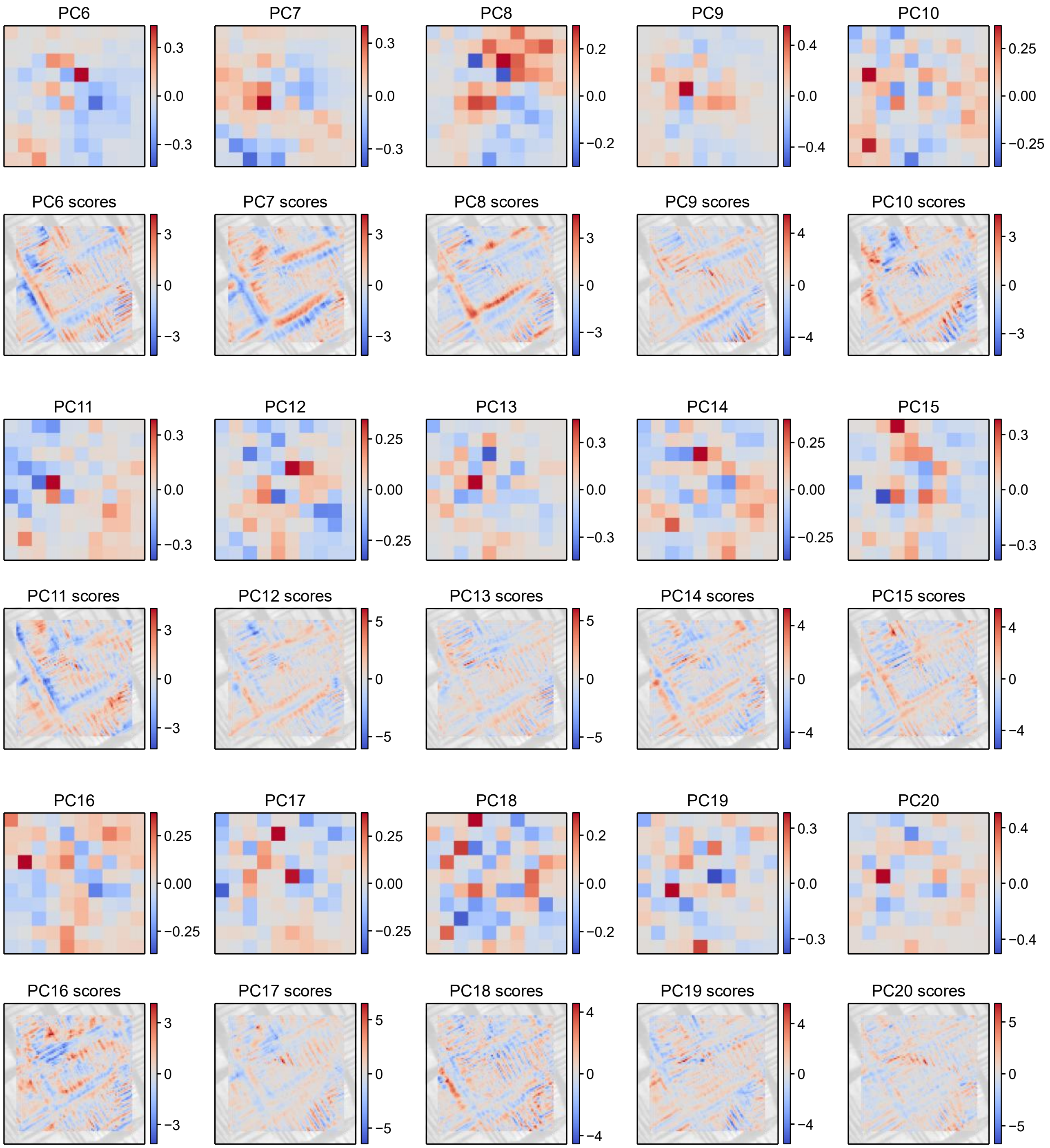

**Figure S2. Additional SHAP–PCA components for PFM–BEPS.** Alternating rows show loading maps and corresponding spatial scores for PC6–PC20. PCA fitting and score standardization use all 6,724 eligible attribution descriptors, as in Figure 2. Percentages report explained variance. Red and blue indicate positive and negative PCA values; each panel has its own symmetric color scale. The grayscale border is ineligible for a complete patch. Maps include training and test locations.

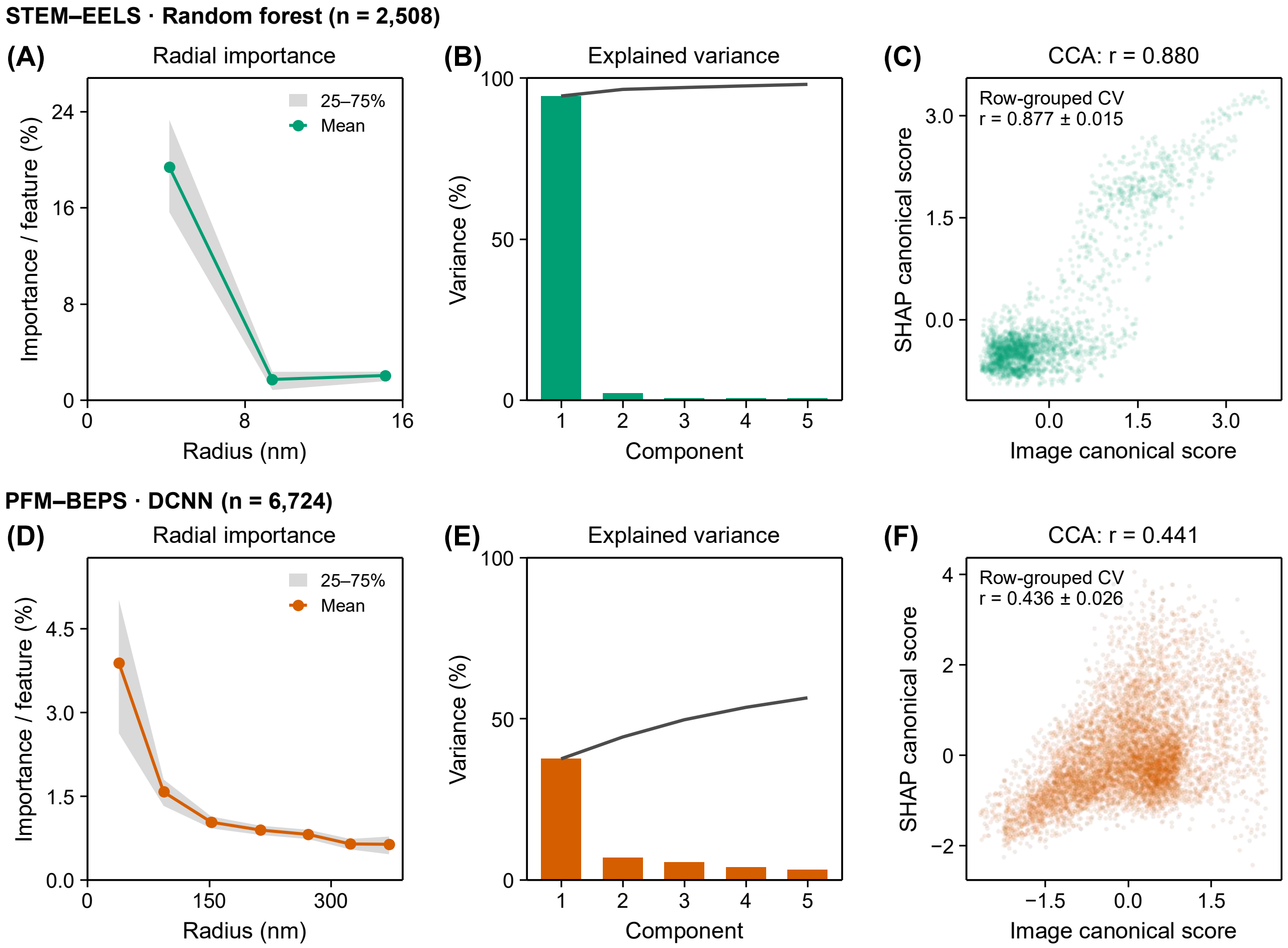

**Figure S3. Additional attribution diagnostics (A–C)** STEM–EELS random-forest results for 2,508 eligible patches. **(D–F)** PFM–BEPS DCNN results for 6,724 eligible patches. **(A, D)** Mean radial importance per pooled feature, with the 25th–75th percentile across patches shaded. Each patch is normalized to 100% total importance before radial averaging. **(B, E)** Individual and cumulative explained variance of the first five attribution PCs. **(C, F)** First image–SHAP canonical score pair fitted to all eligible descriptors. Titles report full-data Pearson correlations; annotations report mean ± sample standard deviation of held-out correlations across five row-grouped folds, with PCA and CCA fitted separately within each fold. Overlapping neighborhoods remain possible across folds. Note: These associations describe predictor attribution rather than causality or independent-specimen generalization.

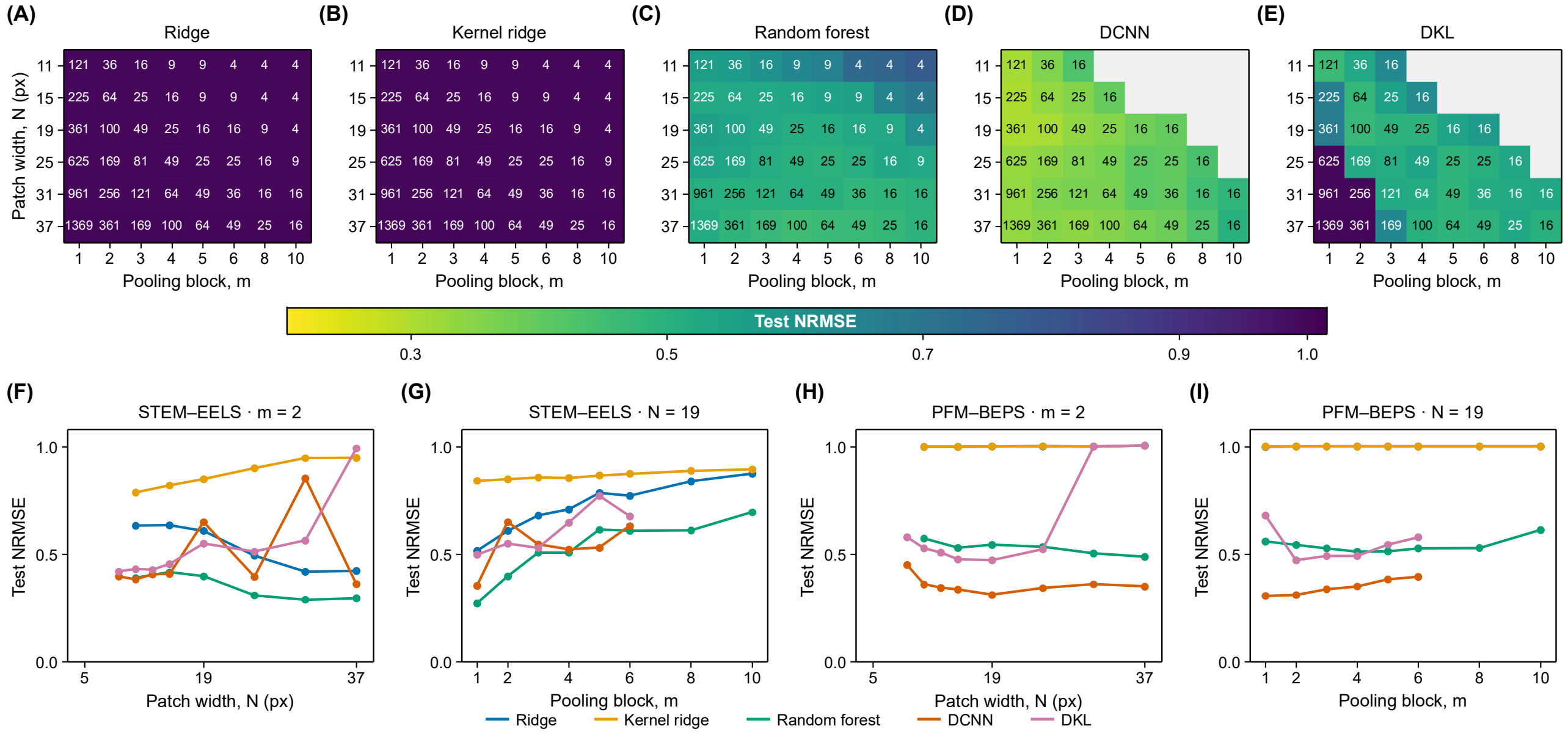

**Figure S4. Additional uniform-compression results**. (A–E) PFM–BEPS error grids for Ridge, kernel Ridge, Random Forest, DCNN, and DKL; colors use the NRMSE scale in Figure 3. (F, G) STEM–EELS error versus patch size at compression factor 2 and versus compression at patch width 19. (H, I) Corresponding PFM–BEPS curves. Cell numbers report retained feature counts.

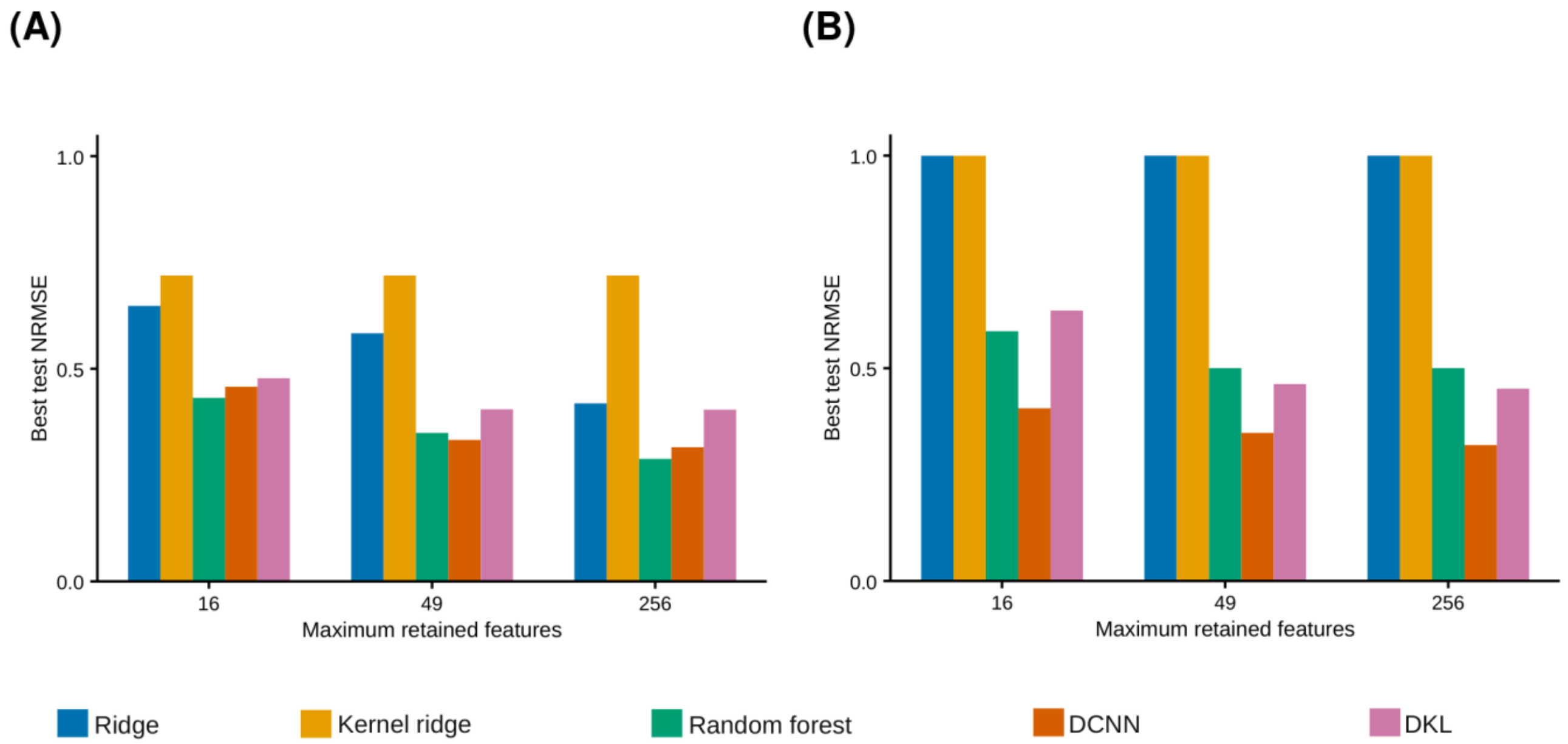

**Figure S5. Best reported test NRMSE** within selected feature budgets for (A) STEM–EELS and (B) PFM–BEPS. Budgets of 16, 49, and 256 retained features are compared using the same model colors as Figure 4. Each bar shows the lowest reported test NRMSE among evaluated configurations with retained feature count no greater than the stated budget. These are descriptive test-error minima rather than validation-selected estimates.

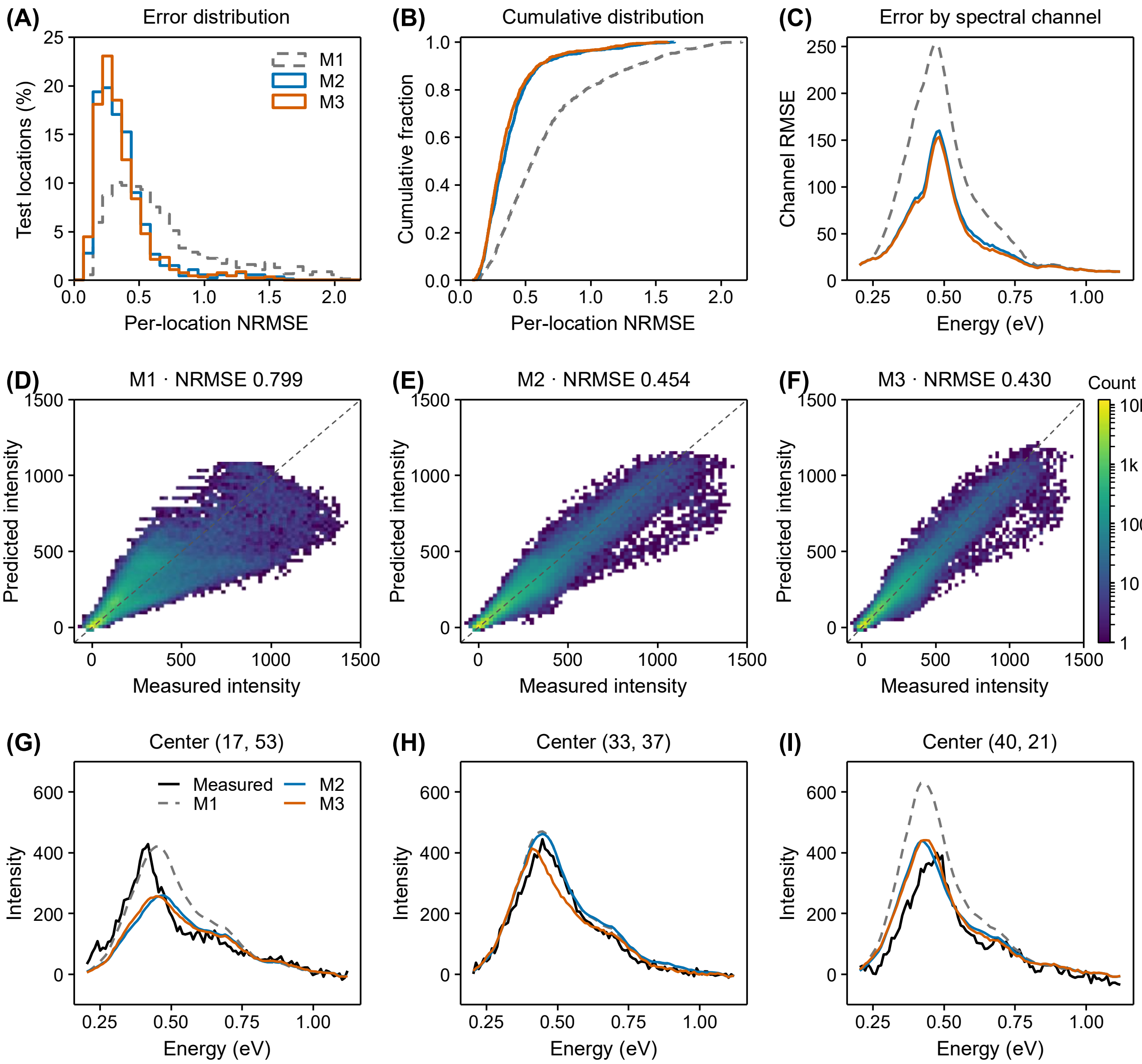

**Figure S6. Held-out diagnostics for Cartesian descriptors in Figure 5.** M1–M3 contain 1, 6, and 27 values, respectively. All panels use 945 held-out centers. (A) Per-location full-spectrum NRMSE distributions. (B) Corresponding cumulative distributions. (C) RMSE by spectral channel. (D–F) Measured versus predicted intensities for M1–M3; colors indicate bin counts on a logarithmic scale, and dashed lines indicate exact agreement. (G–I) Measured and predicted spectra at three shared test centers selected independently of prediction error. Error normalization follows the Supplementary Methods.

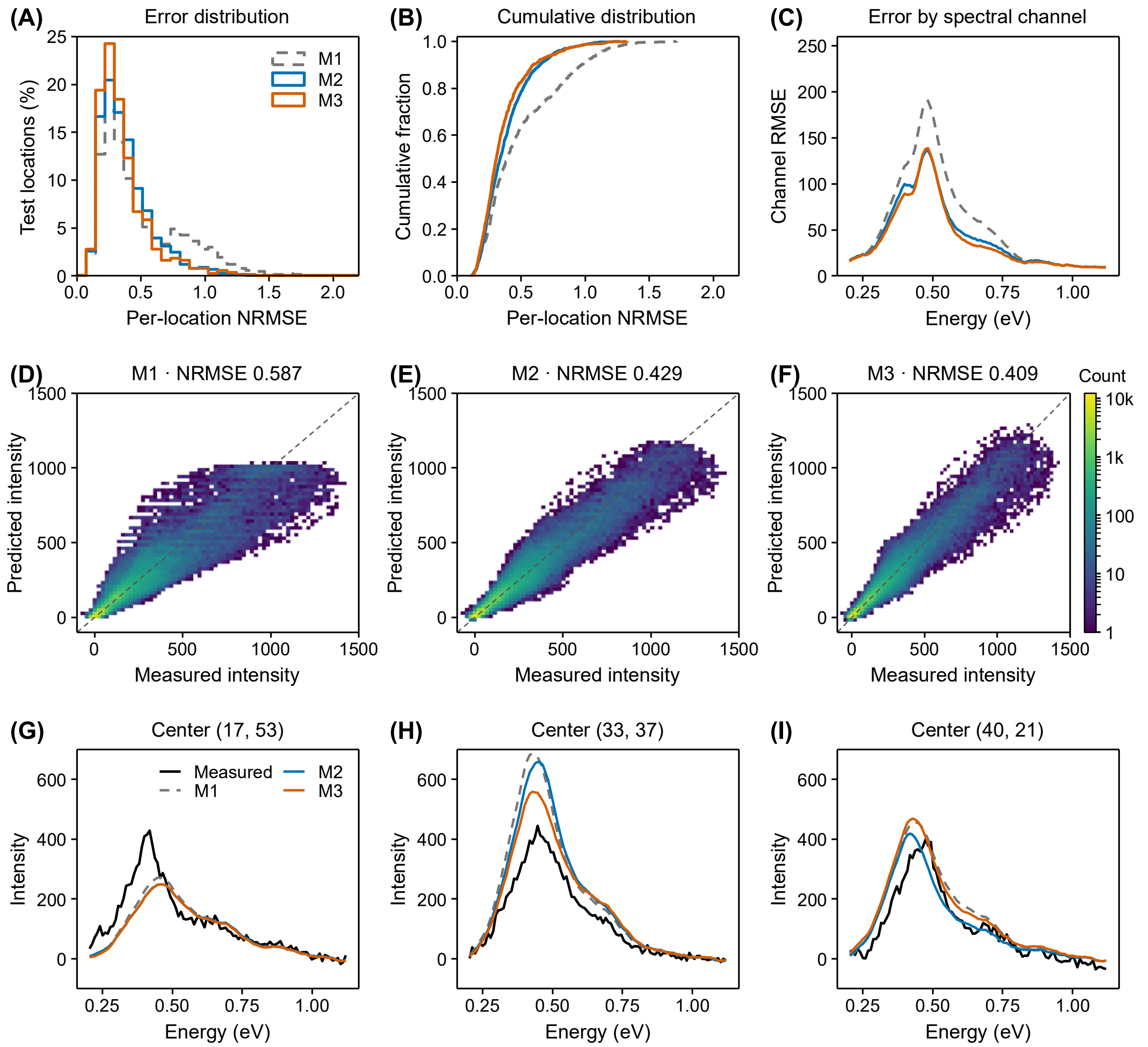

**Figure S7. Held-out diagnostics for polar descriptors in Figure 6.** M1–M3 contain 1, 4, and 15 values, respectively. All panels use the same 945 held-out centers as Figure S6. (A) Per-location full-spectrum NRMSE distributions. (B) Corresponding cumulative distributions. (C) RMSE by spectral channel. (D–F) Measured versus predicted intensities for M1–M3; colors indicate bin counts on a logarithmic scale, and dashed lines indicate exact agreement. (G–I) Measured and predicted spectra at the same three test centers as Figure S6. Error normalization and diagnostic scales match Figure S6.

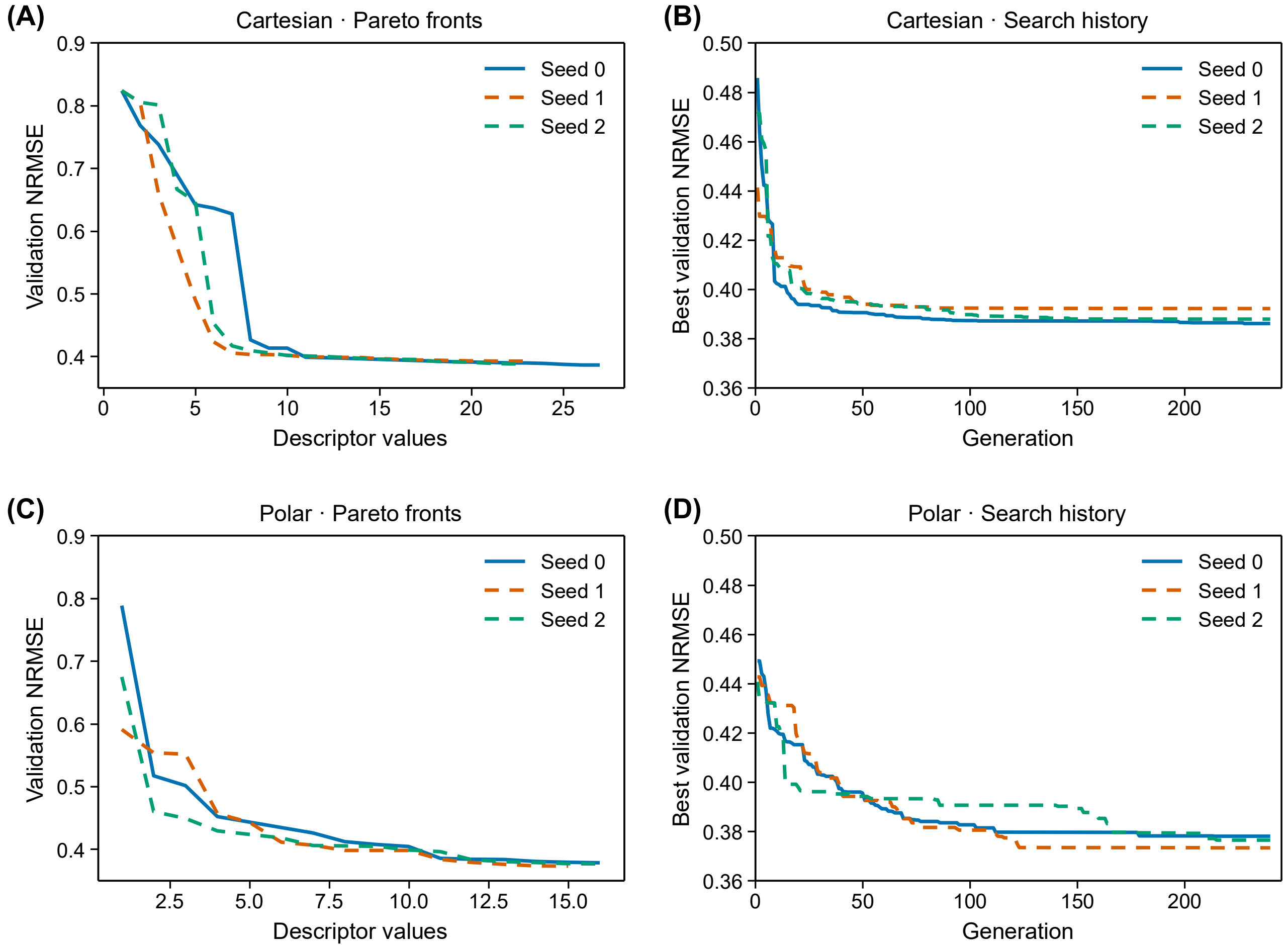

**Figure S8. Genetic-search repeatability**. Cartesian results are shown above polar results. **(A, C)** nondominated fronts from three seeds. **(B, D)**: best validation NRMSE versus generation, shown only through each run's recorded final generation. All runs use fixed data and Ridge alpha = 10. Variation across seeds characterizes optimization repeatability, not independent-image uncertainty.

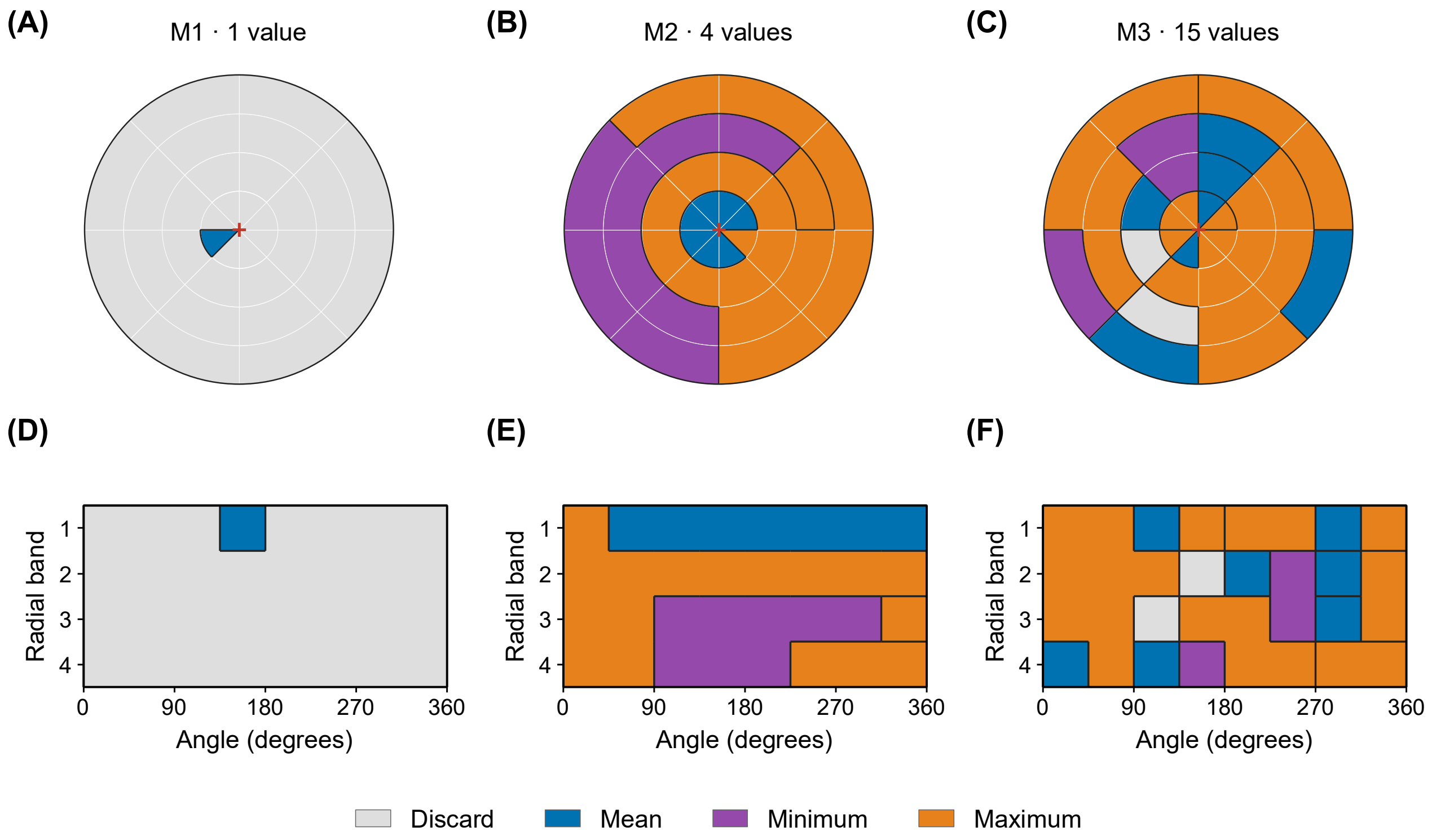

**Figure S9. Polar descriptors and their unwrapped radial–angular representations**. The three candidates contain 1, 4, and 15 retained values. **(A–C)**: masks in image coordinates. **(D–F)**: the corresponding radial-band and angular-bin representations. Gray denotes discarded regions; the other colors denote mean, minimum, and maximum pooling. M1–M3 correspond to Figure 6. Black boundaries separate regions contributing distinct descriptor values, including adjacent regions with the same pooling operation. Angular adjacency wraps across the unwrapped representation's 0°/360° seam.